%% file: prd.tex
\documentclass[aps,prd,twocolumn,superscriptaddress,showpacs,floatfix]{revtex4}
\usepackage{amsmath}
\usepackage{amssymb}
\usepackage{color}
\usepackage{dcolumn}
\usepackage{graphicx}
\usepackage{multirow}
\usepackage{rotating}
\usepackage{bm}
\usepackage{xspace}
\usepackage{lscape}
\usepackage{comment}
\usepackage{color}
\usepackage{slashed}

\newcommand{\pt} {\mbox{$p_T$}}
\newcommand{\dzero}     {D0}
\newcommand{\ttbar}     {\mbox{$t\bar{t}$}\xspace}
\newcommand{\ppbar}     {\mbox{$p\bar{p}$}\xspace}
\newcommand{\ljets}     {\mbox{$\ell$+jets}\xspace}
\newcommand{\mpt}	{\mbox{$\slashed{p}_{T}$}\xspace}
\newcommand{\pythia}    {\mbox{\textsc{pythia}}}
\newcommand{\herwig}    {\mbox{\textsc{herwig}}}
\newcommand{\geant}     {{\sc{geant}}}
\newcommand{\alpgen}    {\mbox{\textsc{alpgen}}}
\newcommand{\mcatnlo}   {\mbox{\textsc{mc@nlo}}}

\newcommand{\ifb}       {fb$^{-1}$}
\newcommand{\tevcent}	{7.60}
\newcommand{\tevtoterr}	{0.41}

\begin{document}

\input title_prd


\section{Introduction}
\label{sec:Introduction}
\input{intro}

\section{CDF and \dzero\ combinations}
\label{sec:inputs}
\input{inputs}

\section{Overview of systematic uncertainties}
\label{sec:combi}
\input{combi}

\section{Results} 
\label{sec:results}
\input{results}

\section{Conclusion}
\label{sec:conclusion}
\input{conclusion}

\section*{Acknowledgments}
\input{acknowledgement.tex}
\input{references}
\end{document}

%% file: title_prd.tex
\hspace{5.2in} \mbox{FERMILAB-PUB-13-432-E}


\title{\boldmath \bf Combination of measurements of the top-quark pair production cross section from the Tevatron Collider}

\input author.tex

\date{September 27th, 2013}

\begin{abstract}
We combine six measurements of the inclusive top-quark pair (\ttbar)
production cross section ($\sigma_{t\bar{t}}$) from data collected
with the CDF and \dzero\ detectors at the Fermilab Tevatron with
proton anti-proton collisions at $\sqrt{s}$ = 1.96~TeV.  The data
correspond to integrated luminosities of up to 8.8~\ifb.  We obtain a
value of $\sigma_{t\bar{t}} = \tevcent \pm \tevtoterr~\mathrm{pb}$ for
a top-quark mass of $m_t=172.5$~GeV.  The contributions to the
uncertainty are 0.20~pb from statistical sources, 0.29~pb from
systematic sources, and 0.21~pb from the uncertainty on the integrated
luminosity.  The result is in good agreement with the
standard model expectation of $7.35^{+0.28}_{-0.33}$\,pb at NNLO+NNLL in pertubative QCD.

\end{abstract}


\maketitle


%% file: author.tex
\affiliation{Institute of Physics, Academia Sinica, Taipei, Taiwan 11529, Republic of China}
\affiliation{Argonne National Laboratory, Argonne, Illinois 60439, USA}
\affiliation{University of Athens, 157 71 Athens, Greece}
\affiliation{Institut de Fisica d'Altes Energies, ICREA, Universitat Autonoma de Barcelona, E-08193, Bellaterra (Barcelona), Spain}
\affiliation{Baylor University, Waco, Texas 76798, USA}
\affiliation{Istituto Nazionale di Fisica Nucleare Bologna, \ensuremath{^{ss}}University of Bologna, I-40127 Bologna, Italy}
\affiliation{University of California, Davis, Davis, California 95616, USA}
\affiliation{University of California, Los Angeles, Los Angeles, California 90024, USA}
\affiliation{Instituto de Fisica de Cantabria, CSIC-University of Cantabria, 39005 Santander, Spain}
\affiliation{Carnegie Mellon University, Pittsburgh, Pennsylvania 15213, USA}
\affiliation{Enrico Fermi Institute, University of Chicago, Chicago, Illinois 60637, USA}
\affiliation{Comenius University, 842 48 Bratislava, Slovakia; Institute of Experimental Physics, 040 01 Kosice, Slovakia}
\affiliation{Joint Institute for Nuclear Research, RU-141980 Dubna, Russia}
\affiliation{Duke University, Durham, North Carolina 27708, USA}
\affiliation{Fermi National Accelerator Laboratory, Batavia, Illinois 60510, USA}
\affiliation{University of Florida, Gainesville, Florida 32611, USA}
\affiliation{Laboratori Nazionali di Frascati, Istituto Nazionale di Fisica Nucleare, I-00044 Frascati, Italy}
\affiliation{University of Geneva, CH-1211 Geneva 4, Switzerland}
\affiliation{Glasgow University, Glasgow G12 8QQ, United Kingdom}
\affiliation{Harvard University, Cambridge, Massachusetts 02138, USA}
\affiliation{Division of High Energy Physics, Department of Physics, University of Helsinki, FIN-00014, Helsinki, Finland; Helsinki Institute of Physics, FIN-00014, Helsinki, Finland}
\affiliation{University of Illinois, Urbana, Illinois 61801, USA}
\affiliation{The Johns Hopkins University, Baltimore, Maryland 21218, USA}
\affiliation{Institut f\"{u}r Experimentelle Kernphysik, Karlsruhe Institute of Technology, D-76131 Karlsruhe, Germany}
\affiliation{Center for High Energy Physics: Kyungpook National University, Daegu 702-701, Korea; Seoul National University, Seoul 151-742, Korea; Sungkyunkwan University, Suwon 440-746, Korea; Korea Institute of Science and Technology Information, Daejeon 305-806, Korea; Chonnam National University, Gwangju 500-757, Korea; Chonbuk National University, Jeonju 561-756, Korea; Ewha Womans University, Seoul, 120-750, Korea}
\affiliation{Ernest Orlando Lawrence Berkeley National Laboratory, Berkeley, California 94720, USA}
\affiliation{University of Liverpool, Liverpool L69 7ZE, United Kingdom}
\affiliation{University College London, London WC1E 6BT, United Kingdom}
\affiliation{Centro de Investigaciones Energeticas Medioambientales y Tecnologicas, E-28040 Madrid, Spain}
\affiliation{Massachusetts Institute of Technology, Cambridge, Massachusetts 02139, USA}
\affiliation{University of Michigan, Ann Arbor, Michigan 48109, USA}
\affiliation{Michigan State University, East Lansing, Michigan 48824, USA}
\affiliation{Institution for Theoretical and Experimental Physics, ITEP, Moscow 117259, Russia}
\affiliation{University of New Mexico, Albuquerque, New Mexico 87131, USA}
\affiliation{The Ohio State University, Columbus, Ohio 43210, USA}
\affiliation{Okayama University, Okayama 700-8530, Japan}
\affiliation{Osaka City University, Osaka 558-8585, Japan}
\affiliation{University of Oxford, Oxford OX1 3RH, United Kingdom}
\affiliation{Istituto Nazionale di Fisica Nucleare, Sezione di Padova, \ensuremath{^{tt}}University of Padova, I-35131 Padova, Italy}
\affiliation{University of Pennsylvania, Philadelphia, Pennsylvania 19104, USA}
\affiliation{Istituto Nazionale di Fisica Nucleare Pisa, \ensuremath{^{uu}}University of Pisa, \ensuremath{^{vv}}University of Siena, \ensuremath{^{ww}}Scuola Normale Superiore, I-56127 Pisa, Italy, \ensuremath{^{xx}}INFN Pavia, I-27100 Pavia, Italy, \ensuremath{^{yy}}University of Pavia, I-27100 Pavia, Italy}
\affiliation{University of Pittsburgh, Pittsburgh, Pennsylvania 15260, USA}
\affiliation{Purdue University, West Lafayette, Indiana 47907, USA}
\affiliation{University of Rochester, Rochester, New York 14627, USA}
\affiliation{The Rockefeller University, New York, New York 10065, USA}
\affiliation{Istituto Nazionale di Fisica Nucleare, Sezione di Roma 1, \ensuremath{^{zz}}Sapienza Universit\`{a} di Roma, I-00185 Roma, Italy}
\affiliation{Mitchell Institute for Fundamental Physics and Astronomy, Texas A\&M University, College Station, Texas 77843, USA}
\affiliation{Istituto Nazionale di Fisica Nucleare Trieste, \ensuremath{^{aaa}}Gruppo Collegato di Udine, \ensuremath{^{bbb}}University of Udine, I-33100 Udine, Italy, \ensuremath{^{ccc}}University of Trieste, I-34127 Trieste, Italy}
\affiliation{University of Tsukuba, Tsukuba, Ibaraki 305, Japan}
\affiliation{Tufts University, Medford, Massachusetts 02155, USA}
\affiliation{University of Virginia, Charlottesville, Virginia 22906, USA}
\affiliation{Waseda University, Tokyo 169, Japan}
\affiliation{Wayne State University, Detroit, Michigan 48201, USA}
\affiliation{University of Wisconsin, Madison, Wisconsin 53706, USA}
\affiliation{Yale University, New Haven, Connecticut 06520, USA}
\affiliation{LAFEX, Centro Brasileiro de Pesquisas F\'{i}sicas, Rio de Janeiro, Brazil}
\affiliation{Universidade do Estado do Rio de Janeiro, Rio de Janeiro, Brazil}
\affiliation{Universidade Federal do ABC, Santo Andr\'{e}, Brazil}
\affiliation{University of Science and Technology of China, Hefei, People's Republic of China}
\affiliation{Universidad de los Andes, Bogot\'{a}, Colombia}
\affiliation{Charles University, Faculty of Mathematics and Physics, Center for Particle Physics, Prague, Czech Republic}
\affiliation{Czech Technical University in Prague, Prague, Czech Republic}
\affiliation{Institute of Physics, Academy of Sciences of the Czech Republic, Prague, Czech Republic}
\affiliation{Universidad San Francisco de Quito, Quito, Ecuador}
\affiliation{LPC, Universit\'{e} Blaise Pascal, CNRS/IN2P3, Clermont, France}
\affiliation{LPSC, Universit\'{e} Joseph Fourier Grenoble 1, CNRS/IN2P3, Institut National Polytechnique de Grenoble, Grenoble, France}
\affiliation{CPPM, Aix-Marseille Universit\'{e}, CNRS/IN2P3, Marseille, France}
\affiliation{LAL, Universit\'{e} Paris-Sud, CNRS/IN2P3, Orsay, France}
\affiliation{LPNHE, Universit\'{e}s Paris VI and VII, CNRS/IN2P3, Paris, France}
\affiliation{CEA, Irfu, SPP, Saclay, France}
\affiliation{IPHC, Universit\'{e} de Strasbourg, CNRS/IN2P3, Strasbourg, France}
\affiliation{IPNL, Universit\'{e} Lyon 1, CNRS/IN2P3, Villeurbanne, France and Universit\'{e} de Lyon, Lyon, France}
\affiliation{III. Physikalisches Institut A, RWTH Aachen University, Aachen, Germany}
\affiliation{Physikalisches Institut, Universit\"{a}t Freiburg, Freiburg, Germany}
\affiliation{II. Physikalisches Institut, Georg-August-Universit\"{a}t G\"{o}ttingen, G\"{o}ttingen, Germany}
\affiliation{Institut f\"{u}r Physik, Universit\"{a}t Mainz, Mainz, Germany}
\affiliation{Ludwig-Maximilians-Universit\"{a}t M\"{u}nchen, M\"{u}nchen, Germany}
\affiliation{Panjab University, Chandigarh, India}
\affiliation{Delhi University, Delhi, India}
\affiliation{Tata Institute of Fundamental Research, Mumbai, India}
\affiliation{University College Dublin, Dublin, Ireland}
\affiliation{Korea Detector Laboratory, Korea University, Seoul, Korea}
\affiliation{CINVESTAV, Mexico City, Mexico}
\affiliation{Nikhef, Science Park, Amsterdam, the Netherlands}
\affiliation{Radboud University Nijmegen, Nijmegen, the Netherlands}
\affiliation{Joint Institute for Nuclear Research, Dubna, Russia}
\affiliation{Institute for Theoretical and Experimental Physics, Moscow, Russia}
\affiliation{Moscow State University, Moscow, Russia}
\affiliation{Institute for High Energy Physics, Protvino, Russia}
\affiliation{Petersburg Nuclear Physics Institute, St. Petersburg, Russia}
\affiliation{Instituci\'{o} Catalana de Recerca i Estudis Avan\c{c}ats (ICREA) and Institut de F\'{i}sica d'Altes Energies (IFAE), Barcelona, Spain}
\affiliation{Uppsala University, Uppsala, Sweden}
\affiliation{Lancaster University, Lancaster LA1 4YB, United Kingdom}
\affiliation{Imperial College London, London SW7 2AZ, United Kingdom}
\affiliation{The University of Manchester, Manchester M13 9PL, United Kingdom}
\affiliation{University of Arizona, Tucson, Arizona 85721, USA}
\affiliation{University of California Riverside, Riverside, California 92521, USA}
\affiliation{Florida State University, Tallahassee, Florida 32306, USA}
\affiliation{Fermi National Accelerator Laboratory, Batavia, Illinois 60510, USA}
\affiliation{University of Illinois at Chicago, Chicago, Illinois 60607, USA}
\affiliation{Northern Illinois University, DeKalb, Illinois 60115, USA}
\affiliation{Northwestern University, Evanston, Illinois 60208, USA}
\affiliation{Indiana University, Bloomington, Indiana 47405, USA}
\affiliation{Purdue University Calumet, Hammond, Indiana 46323, USA}
\affiliation{University of Notre Dame, Notre Dame, Indiana 46556, USA}
\affiliation{Iowa State University, Ames, Iowa 50011, USA}
\affiliation{University of Kansas, Lawrence, Kansas 66045, USA}
\affiliation{Louisiana Tech University, Ruston, Louisiana 71272, USA}
\affiliation{Northeastern University, Boston, Massachusetts 02115, USA}
\affiliation{University of Michigan, Ann Arbor, Michigan 48109, USA}
\affiliation{Michigan State University, East Lansing, Michigan 48824, USA}
\affiliation{University of Mississippi, University, Mississippi 38677, USA}
\affiliation{University of Nebraska, Lincoln, Nebraska 68588, USA}
\affiliation{Rutgers University, Piscataway, New Jersey 08855, USA}
\affiliation{Princeton University, Princeton, New Jersey 08544, USA}
\affiliation{State University of New York, Buffalo, New York 14260, USA}
\affiliation{University of Rochester, Rochester, New York 14627, USA}
\affiliation{State University of New York, Stony Brook, New York 11794, USA}
\affiliation{Brookhaven National Laboratory, Upton, New York 11973, USA}
\affiliation{Langston University, Langston, Oklahoma 73050, USA}
\affiliation{University of Oklahoma, Norman, Oklahoma 73019, USA}
\affiliation{Oklahoma State University, Stillwater, Oklahoma 74078, USA}
\affiliation{Brown University, Providence, Rhode Island 02912, USA}
\affiliation{University of Texas, Arlington, Texas 76019, USA}
\affiliation{Southern Methodist University, Dallas, Texas 75275, USA}
\affiliation{Rice University, Houston, Texas 77005, USA}
\affiliation{University of Virginia, Charlottesville, Virginia 22904, USA}
\affiliation{University of Washington, Seattle, Washington 98195, USA}

\author{T.~Aaltonen\ensuremath{^{\dag}}}
\affiliation{Division of High Energy Physics, Department of Physics, University of Helsinki, FIN-00014, Helsinki, Finland; Helsinki Institute of Physics, FIN-00014, Helsinki, Finland}
\author{V.M.~Abazov\ensuremath{^{\ddag}}}
\affiliation{Joint Institute for Nuclear Research, Dubna, Russia}
\author{B.~Abbott\ensuremath{^{\ddag}}}
\affiliation{University of Oklahoma, Norman, Oklahoma 73019, USA}
\author{B.S.~Acharya\ensuremath{^{\ddag}}}
\affiliation{Tata Institute of Fundamental Research, Mumbai, India}
\author{M.~Adams\ensuremath{^{\ddag}}}
\affiliation{University of Illinois at Chicago, Chicago, Illinois 60607, USA}
\author{T.~Adams\ensuremath{^{\ddag}}}
\affiliation{Florida State University, Tallahassee, Florida 32306, USA}
\author{J.P.~Agnew\ensuremath{^{\ddag}}}
\affiliation{The University of Manchester, Manchester M13 9PL, United Kingdom}
\author{G.D.~Alexeev\ensuremath{^{\ddag}}}
\affiliation{Joint Institute for Nuclear Research, Dubna, Russia}
\author{G.~Alkhazov\ensuremath{^{\ddag}}}
\affiliation{Petersburg Nuclear Physics Institute, St. Petersburg, Russia}
\author{A.~Alton\ensuremath{^{\ddag}}\ensuremath{^{ii}}}
\affiliation{University of Michigan, Ann Arbor, Michigan 48109, USA}
% ...

\collaboration{CDF Collaboration}
\altaffiliation[With visitors from]{
\ensuremath{^{a}}University of British Columbia, Vancouver, BC V6T 1Z1, Canada,
\ensuremath{^{b}}Istituto Nazionale di Fisica Nucleare, Sezione di Cagliari, 09042 Monserrato (Cagliari), Italy,
\ensuremath{^{c}}University of California Irvine, Irvine, CA 92697, USA,
\ensuremath{^{d}}Institute of Physics, Academy of Sciences of the Czech Republic, 182 21, Czech Republic,
\ensuremath{^{e}}CERN, CH-1211 Geneva, Switzerland,
\ensuremath{^{f}}Cornell University, Ithaca, NY 14853, USA,
\ensuremath{^{g}}University of Cyprus, Nicosia CY-1678, Cyprus,
\ensuremath{^{h}}Office of Science, U.S. Department of Energy, Washington, DC 20585, USA,
\ensuremath{^{i}}University College Dublin, Dublin 4, Ireland,
\ensuremath{^{j}}ETH, 8092 Z\"{u}rich, Switzerland,
\ensuremath{^{k}}University of Fukui, Fukui City, Fukui Prefecture, Japan 910-0017,
\ensuremath{^{l}}Universidad Iberoamericana, Lomas de Santa Fe, M\'{e}xico, C.P. 01219, Distrito Federal,
\ensuremath{^{m}}University of Iowa, Iowa City, IA 52242, USA,
\ensuremath{^{n}}Kinki University, Higashi-Osaka City, Japan 577-8502,
\ensuremath{^{o}}Kansas State University, Manhattan, KS 66506, USA,
\ensuremath{^{p}}Brookhaven National Laboratory, Upton, NY 11973, USA,
\ensuremath{^{q}}Queen Mary, University of London, London, E1 4NS, United Kingdom,
\ensuremath{^{r}}University of Melbourne, Victoria 3010, Australia,
\ensuremath{^{s}}Muons, Inc., Batavia, IL 60510, USA,
\ensuremath{^{t}}Nagasaki Institute of Applied Science, Nagasaki 851-0193, Japan,
\ensuremath{^{u}}National Research Nuclear University, Moscow 115409, Russia,
\ensuremath{^{v}}Northwestern University, Evanston, IL 60208, USA,
\ensuremath{^{w}}University of Notre Dame, Notre Dame, IN 46556, USA,
\ensuremath{^{x}}Universidad de Oviedo, E-33007 Oviedo, Spain,
\ensuremath{^{y}}CNRS-IN2P3, Paris, F-75205 France,
\ensuremath{^{z}}Universidad Tecnica Federico Santa Maria, 110v Valparaiso, Chile,
\ensuremath{^{aa}}The University of Jordan, Amman 11942, Jordan,
\ensuremath{^{bb}}Universite catholique de Louvain, 1348 Louvain-La-Neuve, Belgium,
\ensuremath{^{cc}}University of Z\"{u}rich, 8006 Z\"{u}rich, Switzerland,
\ensuremath{^{dd}}Massachusetts General Hospital, Boston, MA 02114 USA,
\ensuremath{^{ee}}Harvard Medical School, Boston, MA 02114 USA,
\ensuremath{^{ff}}Hampton University, Hampton, VA 23668, USA,
\ensuremath{^{gg}}Los Alamos National Laboratory, Los Alamos, NM 87544, USA,
\ensuremath{^{hh}}Universit\`{a} degli Studi di Napoli Federico I, I-80138 Napoli, Italy
}
\noaffiliation
\collaboration{D0 Collaboration}
\altaffiliation[With visitors from]{
\ensuremath{^{ii}}Augustana College, Sioux Falls, SD, USA,
\ensuremath{^{jj}}The University of Liverpool, Liverpool, UK,
\ensuremath{^{kk}}DESY, Hamburg, Germany,
\ensuremath{^{ll}}Universidad Michoacana de San Nicolas de Hidalgo, Morelia, Mexico,
\ensuremath{^{mm}}SLAC, Menlo Park, CA, USA,
\ensuremath{^{nn}}University College London, London, UK,
\ensuremath{^{oo}}Centro de Investigacion en Computacion - IPN, Mexico City, Mexico,
\ensuremath{^{pp}}Universidade Estadual Paulista, S\~{a}o Paulo, Brazil,
\ensuremath{^{qq}}Karlsruher Institut f\"{u}r Technologie (KIT) - Steinbuch Centre for Computing (SCC),
\ensuremath{^{rr}}Office of Science, U.S. Department of Energy, Washington, D.C. 20585, USA
}
\noaffiliation

%% file: intro.tex
The top quark, the most massive elementary particle to date and the
final member of the three families of quarks of the standard model
(SM), was first observed in 1995 by the CDF and \dzero\ experiments at
the Fermilab Tevatron proton-antiproton ($p\bar{p}$)
collider~\cite{top-observation-cdf,top-observation-dzero}.  The large
mass of the top quark of $m_t=173.20 \pm 0.87$~GeV~\cite{tevtopmass}
and its short lifetime of approximately
$10^{-25}$~s~\cite{top-lifetime-expt,top-lifetime-theory} are of
special interest.  The lifetime is far shorter than the hadronization
time, and provides the opportunity to study the properties of
essentially a bare quark. The large mass suggests that the top quark
may play a special role in the mechanism of electroweak symmetry
breaking, and thereby provide sensitivity to probe a broad class of SM
extensions.  In addition, with the recent discovery of a Higgs
boson~\cite{HiggsAtlas, HiggsCMS}, the properties of the top quark are
expected to be related to the stability of the vacuum in the
universe~\cite{vacuum}.

The properties of the top quark can be assessed through precise
determinations of its production mechanisms and decay rates, in
comparison to SM expectations. Particles and couplings
predicted by extensions of the SM can affect the observed production
cross sections of top quarks.  For example, the observed top-quark
pair (\ttbar) production cross section in all of the experimental
final states may be enhanced above the SM expectation by the
production of new resonances~\cite{cdf-ttres,d0-ttres}, or the
observed production cross section in some of the experimental final
states may be altered from the SM expectation by top quark decay into
new channels, such as a hypothesized charged Higgs boson and
$b$~quark~\cite{cdf-H+-searches,d0-H+-searches}.

In this article, we report on the first combination of measurements by
the CDF and \dzero\ experiments at the Fermilab Tevatron of the
inclusive $t\bar{t}$\ production cross section ($\sigma_{t\bar{t}}$), with the
goal of reducing the experimental uncertainty and thereby providing a
better test of the SM prediction.  
The inclusive $t\bar{t}$\ cross section has also been measured at the LHC 
at different center of mass energies~\cite{ATLAS:2012aa,Chatrchyan:2012bra}.
In the remainder of this section, the status of the theoretical
predictions and the experimental signatures of the \ttbar\ final
states are described.
Section~\ref{sec:inputs} reviews all six  measurements, reports the first
combination of the four CDF results, and reviews the combination of the
two D0 results~\cite{d0xsdil}.
The categories of systematic uncertainties and
their correlations among the measurements are detailed in
Sec.~\ref{sec:combi}.  The first combination of the CDF and \dzero\
measurements is reported in Sec.~\ref{sec:results} and conclusions
are drawn in Sec.~\ref{sec:conclusion}.

\subsection{Predictions for the \boldmath $t\bar{t}$ production cross section}

According to the SM, production of top quarks at hadron colliders
takes place through strong interactions that produce \ttbar, or
through electroweak processes that produce a single top quark. At the
Tevatron $p\bar{p}$ collider, with a center of mass energy of
$\sqrt{s}=1.96$~TeV, top quark production occurs mainly through
$t\bar{t}$ production, which is the focus of this article. The
contribution to $t\bar{t}$ production is approximately $85\%$ from
quark-antiquark annihilation ($q\bar{q}\rightarrow t\bar{t}$) and
$15\%$ from gluon-gluon fusion ($gg\rightarrow t\bar{t}$).

SM predictions for inclusive $t\bar{t}$ production at the Tevatron,
calculated to different orders in perturbative quantum chromodynamics
(QCD), are available in
Refs.~\cite{dawson,sm_nlo,sm_nll,sm_nll_new,SMtheory_A, SMtheory_M,
SMtheory_K,xsec_several,SMtheory_NNLO}.  The first calculations at
full next-to-leading-order (NLO) QCD were performed before the
discovery of the top quark~\cite{dawson}, and have been updated using
the more recent CTEQ6.6 parton distribution functions
(PDF)~\cite{sm_nlo}.  These full NLO calculations were further
improved by adding resummations of logarithmic corrections to the
cross section from higher-order soft-gluon radiation, in particular by
including next-to-leading logarithmic (NLL) soft-gluon
resummation~\cite{sm_nll} and the more recent PDF~\cite{sm_nll_new}.
Also available are NLO calculations with soft-gluon resummation to
next-to-next-to-leading logarithmic (NNLL) accuracy, and
approximations at next-to-next-to-leading-order (NNLO) obtained by
re-expanding the result from NLO+NNLL in a fixed-order series in the
strong coupling constant $\alpha_s$ ($\text
{NNLO}_{\text{approx}}$)~\cite{SMtheory_A, SMtheory_M, SMtheory_K,
xsec_several}.  Differences between these calculations include using a
momentum-space or $N$-space approach, and resummation of the total
cross section or integration of the differential cross section over
phase space.

The computation at full NNLO QCD was performed for the first time in
 2013~\cite{SMtheory_NNLO}, with an uncertainty on $\sigma_{t\bar{t}}$
 of approximately 4\% when matched with NNLL soft-gluon
 resummation. To estimate the uncertainty on $\sigma_{t\bar{t}}$ for a
 given top-quark mass, the factorization and renormalization scales
 are changed by factors of two or one half relative to their nominal
 values.  The sensitivity to choice of PDF is evaluated by changing
 all the PDF parameters within their uncertainties~\cite{sm_nlo}.
The predicted values of $\sigma_{t\bar{t}}$ and their corresponding
 uncertainties, calculated with the {\sc Top++} program~\cite{top++}, are
 provided in Table~\ref{tab:theory_predictions} at NLO, NLO+NLL, and
 NNLO+NNLL.  The top-quark mass for these calculations is set to
 $m_t=172.5$~GeV, with the PDF set corresponding to either
 MSTW2008nlo68cl or MSTW2008nnlo68cl~\cite{mstw}.

We use the full NNLO+NNLL prediction as the default value for
 comparison with the measurements since it has the smallest
 uncertainty. The benefit of the recent theoretical advance to full
 NNLO+NNLL consists of an approximate 4\% increase of the cross
 section and a reduction in the scale uncertainty with
 respect to the NLO+NLL prediction.

\begin{table}[htb]
\caption{\label{tab:theory_predictions} SM predictions
of $\sigma_{t\bar{t}}$ at different orders in perturbative QCD, using
{\sc Top++}~\cite{top++}.  }
\begin{tabular}{lccc}
\hline
\hline
Calculation & $\sigma_{t\bar{t}}$ (pb) & $\Delta \sigma_{\rm scale}$ (pb) & $\Delta \sigma_{\rm PDF}$ (pb) \\ \hline \\[-8pt]
NLO		& 6.85 & $^{+0.37}_{-0.77}$ & $^{+0.19}_{-0.13}$  \\[2pt] \\[-10pt]
NLO+NLL		& 7.09 & $^{+0.28}_{-0.51}$ & $^{+0.19}_{-0.13}$  \\[2pt] \\[-10pt]
NNLO+NNLL	& 7.35 & $^{+0.11}_{-0.21}$ & $^{+0.17}_{-0.12}$  \\[2pt]
\hline
\hline
\end{tabular}
\end{table}

\subsection{Experimental final states}

In the SM, the top quark decays through the weak interaction into a
$W^{+}$~boson and a down-type quark, where decays into $W^+s$ and $W^+d$ are expected
to be suppressed relative to $W^+b$ by the square of the
Cabibbo-Kobayashi-Maskawa~\cite{PDG2010} matrix elements $V_{ts}$ and
$V_{td}$. Hence, the decay $t\rightarrow W^{+}b$, and its charge
conjugate, is expected to occur with a branching fraction above
$\approx 99.8\%$.  The $W^{+}$~boson subsequently decays either leptonically
into $e^{+}
\nu_e$, $\mu^{+} {\nu}_\mu$, or $\tau^{+} {\nu}_{\tau}$; or into
$u\bar{d}$ or $c\bar{s}$ quarks~\cite{conjugate}.  All the quarks in
the final state evolve into jets of hadrons.  In studies of $t\bar{t}
\to W^{+}b W^{-}\bar{b}$, different final states are defined by the
decays of the two $W$~bosons.  The main channels are the following:
\begin{description}
\item[(i)] {\bf Dilepton:} Events where both $W$~bosons decay into
  $e {\nu}_e$, $\mu {\nu}_\mu$, or $\tau {\nu}_{\tau}$ with the $\tau$
  decaying leptonically, comprise the dilepton channel. While the
  branching fraction for this channel is only about 4\%, its
  analysis benefits from having a low background.  
\item[(ii)] {\bf Lepton+jets:} This channel (\ljets)
  consists of events where one $W$~boson decays into quarks and the
  other into $e {\nu}_e$, $\mu {\nu}_\mu$, or $\tau {\nu}_{\tau}$ with
  the $\tau$ decaying leptonically. The branching fraction of this
  channel is approximately 35\%.  The main background
  contribution is from the production of $W$ bosons in association with jets. 
\item[(iii)] {\bf All-jets:} Events where both $W$~bosons decay 
 into quarks form the channel with the largest branching fraction of
 about 46\%.  Experimentally, this channel suffers from a large
 background contribution from multijet production.
\end{description}
The remaining two channels are from final states where at least one of
the $W$ bosons decays into $\tau {\nu}_{\tau}$ with the $\tau$ decaying
into hadrons and $\nu_{\tau}$.  These channels have larger
uncertainties, because of the difficulty of reconstructing the
hadronic $\tau$ decays. Hence, they are not used in this combination.

\subsection{Selection and modeling}

The measurement of the $t\bar{t}$ production cross section in each
individual final state requires specific event-selection criteria to
enrich the $t\bar{t}$ content of each sample, a detailed understanding
of background contributions, as well as good modeling of the SM
expectation for the signal and for all background processes. This
section briefly discusses these essential ingredients.  The CDF II
and \dzero\ detectors are described in Refs.~\cite{detector-cdf}
and~\cite{detector-dzero}, respectively.

\subsubsection{Event selection}

Candidate $t\bar{t}$ events are collected by triggering on leptons of
large transverse momentum ($p_T$), and on characteristics of
\ljets or multijet events that depend on the specific final
state.  Differences in topology and kinematic properties between
$t\bar{t}$ events and background processes in each final state are
exploited to enrich the $t\bar{t}$ content of the data samples.  The
discriminating observables used at CDF and \dzero\ for these
selections are based on the properties of jets, electrons, muons, and
the imbalance in transverse momentum (\mpt) in such events. At \dzero,
jets are reconstructed using an iterative midpoint jet cone
algorithm~\cite{d0jets} with ${\cal R}=0.5$ \cite{DeltaR}, while CDF
uses a similar algorithm~\cite{cdfjets} with ${\cal R}=0.4$.
Electrons are reconstructed using information from the electromagnetic
calorimeter, and also require a track from the central tracker that is
matched to the energy cluster in the calorimeter. Muons are
reconstructed using information from the muon system, and also require
a matching track from the central tracker.  Isolation criteria are
applied to identify electrons and muons from $W \to \ell {\nu_\ell}$
decays.  The reconstructed primary interaction vertex must be within
60~cm of the longitudinal center of the detector, corresponding to
about 95\% of the luminous region.

A common feature of all $t\bar{t}$ events are the two $b$-quark jets
from $t \to W b$ decays. The $t\bar{t}$ content of the selected event
samples can therefore be enriched by demanding that they contain
identified $b$~jets.  At CDF, $b$~jets are identified through the
presence of a displaced, secondary vertex~\cite{cdfbid}, while
at \dzero, a neural-network (NN) based $b$-jet identification
algorithm is used for this purpose~\cite{d0bid}.  The NN-based
algorithm combines the information about the impact parameters of
charged particle tracks and the properties of reconstructed secondary
vertices into a single discriminant.

The \mpt\ is reconstructed using the energy deposited in calorimeter
cells, incorporating corrections for the $p_T$ of leptons and
jets. More details on identification criteria for these quantities at
CDF and {\dzero} can be found in Ref.~\cite{fredreview}.

General topologies of each of the three channels are described below,
with specific selections described in the respective references to the
individual measurements cited in Sec.~\ref{sec:inputs}. The selections are designed
so that the channels are mutually exclusive.

\begin{description}
\item[(i)] {\bf Dilepton} candidates are selected by requiring at least two central jets with
high $p_T$; two high-$p_T$, isolated leptons of opposite charge; and
large \mpt to account for the undetected neutrinos from the $W \to
\ell {\nu_\ell}$ decays. Other selections based on the global
properties of the event are applied to reduce backgrounds in each of
the $e^{+}e^{-}$, $e^{\pm}\mu^{\mp}$, and $\mu^{+}\mu^{-}$ final
states.

\item[(ii)] {\bf \ljets} candidates must have at
least three high-$p_T$ jets; one high-$p_T$, isolated electron or muon
within a fiducial region; and significant \mpt\ to account for the
undetected neutrino from the $W \to \ell {\nu_\ell}$ decay. In
addition, requirements are applied on the azimuthal angle between
the lepton direction and \mpt, to reduce contributions from multijet
background.

\item[(iii)] {\bf All-jets} candidates must have at least six central jets with large $p_T$. Events
containing an isolated electron or muon are vetoed, and the event
\mpt\ has to be compatible with its resolution as there are no neutrinos from the $W$ boson decays in this final state.
\end{description}

\subsubsection{Modeling of signal}

A top-quark mass of $m_t=172.5$~GeV, which is close to the measured
top-quark mass~\cite{tevtopmass}, is used in the simulation of \ttbar
production.  Several other values are also simulated in order to
describe the dependence of the measurement of $\sigma_{t\bar{t}}$ on
the assumed value of $m_t$ in the simulation.  This dependence is described in
Sec.~\ref{sec:results}.  

All of the experimental measurements use LO simulations to predict the
fraction of \ttbar\ production passing the selection requirements and
to model the kinematic properties of \ttbar production.  These quantities are
less sensitive to higher-order QCD corrections than the absolute rate.
Contributions to the systematic uncertainty from \ttbar modeling are estimated 
to be approximately 1\% due to the effect of higher-order QCD corrections.
At \dzero, $t\bar{t}$ production and decay are simulated
using the {\alpgen} program~\cite{Alpgen}.  Parton
showering and hadronization are simulated using the
{\pythia}~\cite{Pythia} program. Double-counting of partonic event
configurations is avoided by using a jet-parton matching
scheme~\cite{matching}. The generated events are subsequently
processed through a
\geant-based~\cite{geant} simulation of the
\dzero\ detector.  The presence of additional $p\bar{p}$ interactions
is modeled by overlaying data from random $p\bar{p}$ crossings on the
events.  At CDF, $t\bar{t}$ events are simulated using standalone
{\pythia}, and subsequently processed through a \geant-based
simulation of the CDF II detector~\cite{cdfgeant1, cdfgeant2}, with
additional $p\bar{p}$ collisions modeled using simulation.  Finally,
the events are reconstructed with the same algorithms as used for
data.  Both collaborations implement additional correction factors to
take into account any differences between data and simulation. In
particular, corrections are made to the jet-energy scale,
jet-energy resolution, electron and muon energy scales, trigger
efficiencies, and $b$-jet identification
performance~\cite{dilepton,annsvx,allhad,d0xsdil,d0xsljets}.  At
\dzero, the CTEQ6L1 PDF set is used for event
generation~\cite{pdfd0}, while CDF uses the CTEQ6.6~\cite{sm_nlo} or
CTEQ5L PDF parametrizations~\cite{cteq5}.

\subsubsection{Modeling of backgrounds}

Different sources of backgrounds contribute to different final
states. In the dilepton channel, the dominant source of background is
from Drell-Yan production of $Z$~bosons or virtual photons through
$q\bar{q} \rightarrow Z$ or $\gamma^{*}$ and associated jets, with the
$Z/\gamma^{*}$ decaying into a pair of leptons. In addition,
electroweak diboson production ($WW$, $WZ$, and $ZZ$) and instrumental
background arising from multijet and $W$+jets production, where a jet
is misidentified as a lepton, contribute to the dilepton final
state. At CDF, $W\gamma$ production is considered separately, while at
\dzero\ this contribution is included in the instrumental
background when the $\gamma$ is misidentified as a lepton or as a jet.
For \ljets final states, the major background contribution is from
$W$+jets production, where the $W$ boson decays into $\ell
{\nu_\ell}$.  Backgrounds from single top-quark production, diboson
production, $Z/\gamma^*$+jets, and multijet production are also
considered.  The dominant background contribution to all-jets events
is from multijet production processes.

Contributions from $Z/\gamma^*$+jets and $W$+jets backgrounds are
modeled using \alpgen, followed by \pythia\ for parton showering and
hadronization. Contributions from heavy flavor (HF) quarks, namely
from $W+b\bar{b}$, $W+c\bar{c}$, $W+c$, $Z/\gamma^{*}+b\bar{b}$, and
$Z/\gamma^{*}+c\bar{c}$ are simulated separately.

The diboson contributions to dilepton and \ljets final states are
simulated using standalone \pythia, normalized to the NLO cross
section calculated using \textsc{mcfm}~\cite{mcfm}. Single top-quark
contributions are simulated using the {\textsc{comphep}}
generator~\cite{CompHep} at \dzero, and \textsc{madevent}
~\cite{madevent} at CDF, and normalized to the approximate
NNNLO~\cite{singletopd0} and NLO~\cite{singletopcdf} predictions,
respectively.  The separate background contribution from $W\gamma$
production at CDF is simulated using the \textsc{baur} program~\cite{baurmc}.

The instrumental and multijet backgrounds are estimated using
data-driven methods in different ways for each final state
at CDF and \dzero.

%% file: inputs.tex
We present the first combination of four CDF measurements, which gives
the most precise CDF result to date, and then review the result of a
published combination of two \dzero\ measurements.

\subsection{CDF measurements and their combination}

CDF includes four measurements in the combination: one from the
dilepton channel~\cite{dilepton}, two from the lepton+jets
channel~\cite{annsvx}, and one from the all-jets channel~\cite{allhad}.
Table~\ref{tab:cdfcombi} summarizes these CDF measurements of
$\sigma_{t\bar{t}}$ and their uncertainties.  A detailed description
of the sources of systematic uncertainty and their correlations is
given in Section~\ref{sec:combi}.

\input{table_cdfcombi.tex}

The dilepton (DIL) measurement, $\sigma_{t\bar{t}}= 7.09 \pm 0.83$~pb,
relies on counting events with at least one identified $b$~jet, and
uses the full Run II data set corresponding to an integrated luminosity of
8.8~\ifb~\cite{dilepton}. Backgrounds from diboson and $Z/\gamma^*$
events are predicted from simulation, with additional correction
factors extracted from control samples in data.  The largest
systematic uncertainties for this measurement are from the luminosity
and the modeling of the detector's $b$-jet identification.

The two CDF measurements in the \ljets channel are based on 4.6~\ifb\
of data and apply complementary methods to discriminate signal from
background~\cite{annsvx}.  The first measurement, $\sigma_{t\bar{t}}=
7.82 \pm 0.56$~pb, uses an artificial neural network to exploit
differences between the kinematic properties of signal and $W$+jets
background, without employing $b$-jet identification.  This analysis
is referred to as LJ-ANN.  Due to the large mass of the top quark, its
decay products have larger $p_T$ and are more isotropic than the main
backgrounds from $W$+jets and multijet production. Seven kinematic
properties are selected for analysis in an artificial NN in order to
minimize the statistical uncertainty and the systematic uncertainty
from the calibration of the jet energy.  Since $W$+jets production is
the dominant background in the \ljets channel before the application
of $b$-jet identification requirements, the NN is trained using only
\ttbar and $W$+jets simulated samples. The number of
\ttbar events is then extracted from a maximum likelihood fit 
to the distribution of NN output in data with three or more jets. The
largest systematic uncertainties are from the calibration of jet
energy and the modeling of the \ttbar signal.

The second \ljets measurement, $\sigma_{t\bar{t}}= 7.32 \pm 0.71$~pb,
suppresses the dominant $W$+jets background by reconstruction of
displaced secondary vertices to identify $b$~jets. This analysis is
referred to as LJ-SVX.  The $\sigma_{t\bar{t}}$ is extracted from a
maximum likelihood fit to the observed number of events in data with
at least one identified $b$~jet, given the predicted background.  The
$W$+HF contribution is determined by applying the $b$-jet
identification efficiency and a corrected HF fraction to an estimate
of $W$+jets before $b$-jet identification.  The HF fraction predicted
by the simulation is an underestimate of the yield in data, and a
correction factor is derived from a data control sample. The estimate
of $W$+jets is the number of observed events in data before $b$-jet
identification minus the contribution from other processes (\ttbar,
multijet, single-top, diboson, and $Z/\gamma^{*}$+jets).  The
contribution from events with jets misidentified as $b$~jets is found
by applying a parameterized probability function to the data before the
$b$-jet identification requirement.  The largest systematic
uncertainties in this method arise from the correction for the $W$+HF
background and the modeling of the $b$-jet identification efficiency.

Both \ljets measurements reduce the uncertainty from luminosity by
using the ratio of the \ttbar to the $Z/\gamma^*$ cross sections
measured concurrently. This ratio is multiplied by the more precise
theoretical prediction for the $Z/\gamma^*$ cross
section~\cite{cdf-zxs}, thereby replacing the 6\% uncertainty on
luminosity with a 2\% uncertainty from the smaller theoretical and
experimental uncertainties on the $Z/\gamma^*$ cross section.

The \ljets measurements use subsets of events that pass a common
selection. Their 32\% statistical correlation is evaluated through 1000
simulated experiments.  The $\sigma_{t\bar{t}}$ is extracted for each
such simulated experiment; for LJ-ANN, through a maximum likelihood
fit to the NN distribution, and for LJ-SVX, through the observed
number of events with at least one identified $b$~jet in each
simulated experiment. 

In the all-jets (HAD) measurement, $\sigma_{t\bar{t}}= 7.21 \pm
1.28$~pb, a signal sample is selected by requiring six to eight jets in
an event~\cite{allhad}.  Additional criteria require the presence of
identified $b$~jets and restrictions on the value of a NN
discriminant.  The latter involves 13 observables as input, and is
trained to suppress the large backgrounds from multijet events.  To
improve the statistical significance of the measurement, the
requirement on the value of the discriminant is optimized separately
for events with only one $b$~jet and for events with more than one
$b$~jet.  The $\sigma_{t\bar{t}}$ value is extracted from a
simultaneous fit to the reconstructed top-quark mass in both
samples. The measurement uses only data corresponding to an integrated luminosity of 2.9~\ifb, but the largest
single uncertainty arises from the limited knowledge of the
calibration of jet energy.

To combine the CDF measurements, a best linear unbiased estimate
(BLUE)~\cite{lyons1,lyons2,valassi} is calculated for
$\sigma_{t\bar{t}}$ with the goal of minimizing the total uncertainty.
A covariance matrix is constructed from the statistical and systematic
uncertainties of each result, and from their statistical and
systematic correlations.  The matrix is inverted to obtain a weight
for each result.  These weights are applied to the results to
obtain the best estimate.

Three iterations of the BLUE combination procedure are performed to
eliminate a small bias.  For a measurement with $N$ observed events,
inspection of the simple expression $\sigma_{t\bar{t}}= (N-B) / (\epsilon \mathcal{L})$ shows that the uncertainty on the background
estimate $B$ gives a systematic uncertainty on $\sigma_{t\bar{t}}$
that is independent of the measured value of
$\sigma_{t\bar{t}}$. However, the uncertainties on the
\ttbar\ selection efficiency $\epsilon$ and the luminosity
$\mathcal{L}$ produce systematic uncertainties on $\sigma_{t\bar{t}}$
that are directly proportional to the measured value of
$\sigma_{t\bar{t}}$.  This means that measurements that observe a low
value for $\sigma_{t\bar{t}}$ have a smaller systematic uncertainty
and a larger weight in the BLUE combination than measurements that
observe a high value for $\sigma_{t\bar{t}}$.  Hence, the BLUE
combination underestimates $\sigma_{t\bar{t}}$ and its
uncertainty. This bias is removed by calculating the size of the
systematic uncertainties on $\sigma_{t\bar{t}}$ from the \ttbar\
selection efficiency and luminosity by using the BLUE combination
value from the previous iteration, instead of each measurement's value
of $\sigma_{t\bar{t}}$.  The first iteration uses an arbitrary initial
value of 6~pb.  Simulated experiments show that this procedure removes
the bias.

The combined CDF measurement is 
\begin{displaymath}
\sigma_{t\bar{t}}(\mathrm{CDF}) = 7.63 \pm 0.31 \thinspace {\mathrm{(stat)}} \pm 0.36 \thinspace {\mathrm{(syst)}} 
\pm 0.15 \thinspace {\mathrm{(lumi)}}~\mathrm{pb},
\end{displaymath}
\noindent for $m_t=172.5$~GeV.  The total uncertainty is
0.50~pb.  Table~\ref{tab:cdfcombi} shows the individual
contributions to the uncertainties.  The luminosity uncertainty quoted
above is the sum in quadrature of two sources of uncertainty, from the
inelastic $p\bar{p}$ cross section and from detector-specific effects.
The combination has a $\chi^2$ of 0.86 for three degrees of freedom,
corresponding to a probability of 84\% to have a less consistent set
of measurements.

The largest weight in the BLUE combination of CDF measurements is 70\%
for the \ljets channel LJ-ANN result. The dilepton result has a weight
of 22\%, and the measurement using $b$-jet identification in the
\ljets channel has a weight of 15\%. The measurement in the all-jets
channel has a weight of $-7$\%.  Such negative weights can occur when
the correlation between the two measurements is larger than the ratio
of their total uncertainties~\cite{lyons1}.  The correlation matrix,
including statistical and systematic effects, is given in
Table~\ref{tab:cdfcorrelation}.  The largest correlation is 51\%
between the DIL and HAD measurements, due to the correlation between
systematic uncertainties on detector modeling (primarily $b$-jet
identification), signal modeling, jet energy scale, and luminosity.
Next largest is the 50\% correlation between the LJ-ANN and LJ-SVX
measurements, which arises from a subset of common events and
correlation between systematic uncertainties from signal modeling, jet
energy scale, and normalization of the $Z/\gamma^*$ cross section.
The central value and the total uncertainty change by less than
0.01~pb when the statistical correlation of 32\% between the LJ-ANN and LJ-SVX
measurements is varied by 10\% absolute to 22\% or 42\%.

\begin{table}[htb]
\caption{\label{tab:cdfcorrelation} Correlation matrix for CDF 
  $\sigma_{t\bar{t}}$ measurements, including statistical
and systematic correlations among the methods.}
\begin{tabular}{lccccc}
\hline \hline
Correlation	& LJ-ANN	& LJ-SVX 	& DIL	& HAD \\
\hline
LJ-ANN       	& 1 	& 0.50	&  0.25 & 0.34 \\
LJ-SVX 		& 	& 1 	&  0.44 & 0.47 \\
DIL 		& 	& 	& 1 	& 0.51 \\
HAD		& 	& 	& 	& 1 \\
\hline \hline
\end{tabular}
\end{table}

\subsection{{\dzero} measurements and their combination}

{\dzero} includes two measurements in the combination: one from the
dilepton channel and one from the \ljets channel.  In the dilepton
channel, using data corresponding to an integrated luminosity of
5.4~\ifb, {\dzero} measures $\sigma_{t\bar{t}}= 7.36
^{+0.90}_{-0.79}$~pb through a likelihood fit to a discriminant based
on a NN $b$-jet identification algorithm~\cite{d0xsdil}.  The
$\sigma_{t\bar{t}}$ is extracted from a fit to the distribution of the
smallest of the NN output values from the two jets of highest energy.
The total uncertainty is not limited by the finite sample size but by
the systematic uncertainty on the integrated luminosity.

In the \ljets channel, using data corresponding to an integrated
luminosity of 5.3~\ifb, {\dzero} measures $\sigma_{t\bar{t}} = 7.90
^{+0.78}_{-0.69}$~pb by selecting events with at least three jets and
splitting them into subsamples according to the total number of jets
and the number of identified $b$~jets~\cite{d0xsljets}. In the
background-dominated subsamples (three-jet events with no $b$~jet,
three-jet events with one $b$~jet, and events with at least four jets
and no $b$~jet), a random forest multivariate
discriminant~\cite{TMVA} is used to separate signal from background.  The
$\sigma_{t\bar{t}}$ is extracted by fitting simultaneously the direct
event count in the subsamples with a large $t\bar{t}$ content
(three-jet events with at least two $b$~jets, events with at least
four jets and one $b$~jet, and events with at least four jets and at
least two $b$-jets) and the random forest discriminant in the
background-dominated samples. The leading
systematic uncertainties are treated as nuisance parameters,
constrained by Gaussian prior probability density functions, that are allowed to vary in the fit.
The dominant systematic uncertainty is from the
uncertainty on the luminosity, followed by uncertainties from the
modeling of the detector.

The measurements in the dilepton and \ljets channels have been
combined and published in the dilepton paper~\cite{d0xsdil} with the
same nuisance-parameter technique used in the individual measurements,
accounting for correlations among common systematic sources.  The
result is $\sigma_{t\bar{t}}(\mathrm{\dzero}) = 7.56
^{+0.63}_{-0.56}$~pb. For the combination with CDF, we separate the
statistical and systematic contributions into the categories discussed
in the next section.  We also use the average of the asymmetric
uncertainties of the original {\dzero} measurement. This gives for the combined
{\dzero } measurement
\begin{displaymath}
\sigma_{t\bar{t}}(\mathrm{\dzero}) = 7.56\pm 0.20 {\mathrm{(stat)}}\pm 0.32 \thinspace {\mathrm{(syst)}} \pm 0.46 \thinspace {\mathrm{(lumi)}}~\mathrm{pb},
\end{displaymath}
\noindent for $m_t=172.5$~GeV.  The total uncertainty is
0.59~pb.  Table~\ref{tab:cdfd0} provides the individual contributions
to the uncertainties.  The luminosity uncertainty quoted above is the
sum in quadrature of two sources of uncertainty, from the inelastic
$p\bar{p}$ cross section and from detector-specific effects.

%% file: table_cdfcombi.tex
\begin{table*}[tb]
 \caption{CDF measurements of $\sigma_{t\bar{t}}$ and their combination (in pb), with individual contributions to their uncertainties (in pb).}
 \label{tab:cdfcombi}
\begin{center}
\begin{tabular}{lccccc}
\hline
\hline
 & DIL & LJ-ANN & LJ-SVX & HAD & CDF combined \\ 
 \hline
Central value of $\sigma_{\bar{t}t}$ & 7.09 & 7.82 & 7.32 & 7.21 & 7.63 \\ 
\hline
Sources of systematic uncertainty & & & & & \\
\hline
Modeling of the detector  & 0.39 & 0.11 & 0.34 & 0.41 & 0.17 \\
Modeling of signal  & 0.23 & 0.23 & 0.23 & 0.44 & 0.21 \\
Modeling of jets & 0.23 & 0.23 & 0.29 & 0.71 & 0.21 \\
Method of extracting $\sigma_{t\bar{t}}$  & 0.00 & 0.01 & 0.01 & 0.08 & 0.01 \\
Background modeled from theory & 0.01 & 0.13 & 0.29 & --  & 0.10 \\
Background based on data & 0.15 & 0.07 & 0.11 & 0.59 & 0.08 \\
Normalization of $Z/\gamma^*$ prediction   & -- & 0.16 & 0.15 & -- & 0.13 \\
Luminosity: inelastic $p \bar{p}$ cross section & 0.28 & -- & --  & 0.29 & 0.05 \\
Luminosity: detector & 0.30 & 0.02 & 0.02 & 0.30 & 0.06 \\
\hline
Total systematic uncertainty & 0.67 & 0.41 & 0.61 & 1.18 & 0.39 \\
\hline
Statistical uncertainty & 0.49 & 0.38 & 0.36 & 0.50 & 0.31 \\ 
\hline
 Total uncertainty & 0.83 & 0.56 & 0.71 & 1.28 & 0.50 \\
\hline
\hline
\end{tabular}
 \end{center}
 \end{table*}

%% file: combi.tex
Sources of systematic uncertainty have been categorized into nine
classes with similar correlation properties to facilitate the
combination of the measurements.  Below, we discuss each component of
the uncertainty on the combined cross section.  The values of the CDF
and {\dzero} systematic uncertainties are summarized in
Tables~\ref{tab:cdfcombi} and ~\ref{tab:cdfd0}.

\subsection{Modeling of the detector}

This category includes detector-specific uncertainties on the trigger
and lepton-identification efficiency, $b$-jet identification
efficiency, and modeling of multiple \ppbar\ interactions. In
addition, for CDF measurements, this category includes the uncertainty
on the fraction of the luminous region within the acceptance of the
CDF II detector, track-identification efficiencies for the LJ-ANN and
LJ-SVX measurements, and the uncertainty on the lepton-energy scales.
For {\dzero} measurements, additional uncertainties in this
category arise from vertex reconstruction and identification
efficiency and lepton-energy resolution.  These sources are treated as
correlated within the same experiment, but uncorrelated between experiments.

\subsection{Modeling of signal}
The uncertainties in this category arise from several sources and are
considered fully correlated among all measurements.  

\begin{description}

\item[(i)] {\bf {\boldmath $t \bar{t}$} generator}: This is the source of 
the largest contribution ($1-2\%$) to the signal modeling
systematic. For both CDF and {\dzero} measurements this uncertainty
includes the difference between {\pythia} and {\herwig}~\cite{herwig}
samples resulting from different models for hadronization, for parton showering and for the
underlying event, which describes the remnants of the $p$ and
$\bar{p}$ break-up accompanying the hard partonic
collision. Uncertainties from higher-order QCD corrections are also
included for {\dzero} measurements by comparing results from {\alpgen}
to {\mcatnlo}~\cite{mcatnlo}.  Although there are reported measurements
of a larger-than-expected forward-backward
asymmetry~\cite{cdf-afb,d0-afb} that could be due to non-SM sources,
no additional systematic is assigned as its size would be highly
dependent on the particular hypothesis for the source.  

\item[(ii)] {\bf Parton distribution functions}: The uncertainties on the PDF 
reflect the uncertainty on determining the probability of finding a
particular parton carrying a particular fraction of the $p$ or
$\bar{p}$ momentum. This in turn affects the kinematic distributions
of the final-state particles in \ttbar\ production and decay, as well
as the event selection efficiency.  The default acceptances are
calculated using the LO CTEQ5L and CTEQ6L PDF sets for CDF and {\dzero},
respectively. The systematic uncertainty includes uncertainties
evaluated using the prescribed NLO error vectors from CTEQ6M for CDF,
and CTEQ6.1M for D0, following the recommendations of the CTEQ
collaboration~\cite{cteq61m}. For CDF measurements, this uncertainty
also includes the difference between the central values from LO and
NLO PDF~\cite{topmassprd}.

\item[(iii)] {\bf Initial and final-state radiation}: 
The amount of gluon radiation from partons in the initial or final
state, which affects the \ttbar\ efficiency and kinematic properties,
is set by parameters of the {\pythia} generator used to simulate
\ttbar\ events. The uncertainties on these parameters are taken from a
study of initial state radiation in Drell-Yan events, $q\bar{q} \to
Z/\gamma^* \to
\mu^{+}\mu^{-}$, that share the same initial $q\bar{q}$ state as
most of the \ttbar signal~\cite{isr-fsr-cdf,topmassprd}. 

\item[(iv)] {\bf Color reconnection}: 
This uncertainty is evaluated by comparing {\pythia} configurations with
different parameters that affect the exchange of momentum and energy
via gluons between the color-connected top-quark and antitop-quark
systems. Specifically, the difference in \ttbar\ efficiency obtained
with {\pythia} using the {\sc {\mbox{A-pro}}} and the {\sc
{\mbox{ACR-pro}}} configurations~\cite{skands} is quoted as the systematic
uncertainty~\cite{topmassprd}.

\item[(v)] {\bf Leptonic decay branching fractions for $W$ bosons}: 
This uncertainty alters the proportion of \ttbar\ decays that cause
the dilepton, \ljets, and all-jets final states. It is evaluated by
changing the branching fractions in the $W$-boson decay by their
uncertainties~\cite{PDG2010}.

\end{description}

\subsection{Modeling of jets}
Uncertainties on the modeling of jets affect the \ttbar\ selection
efficiency and the kinematic distributions used to extract
$\sigma_{t\bar{t}}$. They arise from the calibration of light-quark
and $b$-jet energies, and modeling of jet reconstruction and
resolution in the simulation.  These sources are treated as
correlated within each experiment, and uncorrelated between experiments.

\begin{description}

\item[(i)] {\bf Jet-energy scale}: 
This uncertainty arises from uncertainties in calibrating  jet energy
using test-beam data (CDF), as well as $\gamma$+jets and dijet 
events (CDF and {\dzero}). The effect on the measurement is evaluated by 
replacing the jet energies in the nominal simulated samples with  energies 
changed by their estimated systematic uncertainties.

\item[(ii)] {\bf \boldmath{$b$}-jet energy scale}: 
This uncertainty accounts for the difference in energy between jets
originating from light-flavor quarks or gluons, and from
$b$~quarks. For CDF, it includes uncertainties on branching fractions
of semileptonic decays of $b$ and $c$~quarks; uncertainties on
$b$-quark fragmentation; and the uncertainty on the calorimeter
response to $b$ and $c$ hadrons.  For {\dzero}, the sources are
uncertainties on parameters for $b$-quark fragmentation, and the
difference in calorimeter response to jets from $b$ and light
quarks. More details can be found in Ref.~\cite{topmassprd}.

\item[(iii)] {\bf Jet reconstruction and identification}: 
This uncertainty is specific to {\dzero} results, and covers the uncertainty
on correction factors applied to simulation to match the jet
identification efficiency in data, and on factors used to adjust the
jet resolution in simulation to that observed in data.

\end{description}

\subsection{Method for extracting  {\boldmath$\sigma_{t\bar{t}}$}}

This uncertainty is different for each method, and arises from the limited
size of the simulated samples or from the dependence of the calibration on the
specific analysis. It is uncorrelated among all measurements.

\subsection{Background modeled from theory}
For both experiments, this uncertainty includes the uncertainty on the
heavy-flavor fraction in $W$+jets events, uncertainties on the normalization
of the electroweak background (diboson and single top-quark
production), and the dependence on the renormalization and
factorization scale in $W$+jets simulation. Details about specific
modeling of background and the treatment of systematic uncertainties
are in the references for each individual measurement.  Since
these uncertainties are related to the theoretical description of the
background, this source is treated as correlated among all
measurements.

\subsection{Background based on data}
This source covers uncertainties on multijet background in both
experiments; the uncertainty in the modeling of the
$Z/\gamma^*$+jets background, obtained from data by {\dzero}; and
uncertainties on misidentification of jets from charm and
light-flavor quarks as $b$~jets at CDF. This source is considered
uncorrelated among all measurements.

\subsection{Normalization of \boldmath{$Z/\gamma^*$} predictions}
This uncertainty is applicable only to the LJ-ANN and LJ-SVX
measurements by CDF, which exploit the ratio of observed $t\bar{t}$ to
$Z/\gamma^*$ production and therefore involve the normalization using
the predicted $Z/\gamma^*$ cross section~\cite{cdf-zxs}. It includes
the uncertainty on the predicted $Z/\gamma^*$+jets cross section, and
the contributions to the uncertainty on the measured $Z/\gamma^*$+jets
cross section from the background estimate, and from the choice of
renormalization and factorization scale for the $Z/\gamma^*$+jets
simulation.

\subsection{Luminosity uncertainty}

The luminosity uncertainty has two sources:

\begin{description}

\item[(i)] {\bf Inelastic \boldmath{$p \bar{p}$} cross section}: 
The total inelastic $p\bar{p}$ cross section~\cite{inelasticxs} has an
uncertainty of 4.0\%.  This source is correlated among all
measurements but does not affect the CDF LJ-ANN and LJ-SVX
measurements, which exploit the ratio of $t\bar{t}$ to $Z/\gamma^*$
production rates.

\item[(ii)] {\bf Detector-specific luminosity uncertainty}:
This contribution is from detector effects and is approximately
4.5\%~\cite{lumdet}. This source is treated as correlated for
measurements within the same experiment, and uncorrelated between
experiments.  This uncertainty is negligible for CDF LJ-ANN and LJ-SVX
measurements, which exploit the ratio of production rates.

\end{description}

%% file: results.tex
The CDF and {\dzero} $\sigma_{t\bar{t}}$ measurements are combined 
using the BLUE method described in Sec.~\ref{sec:inputs}, with inputs 
from the first two columns of Table~\ref{tab:cdfd0}, yielding the
Tevatron average of
\begin{displaymath}
\sigma_{t\bar{t}} = \tevcent \pm 0.20 \thinspace {\mathrm{(stat)}} \pm 0.29 \thinspace {\mathrm{(syst)}} 
\pm 0.21 \thinspace {\mathrm{(lumi)}}~\mathrm{pb},
\end{displaymath}
\noindent assuming $m_t=172.5$~GeV. The total uncertainty is 0.41~pb. The luminosity uncertainty quoted above is the sum in
quadrature of two sources of uncertainty, from the inelastic
$p\bar{p}$ cross section and from detector-specific effects.  The
individual contributions to the systematic uncertainties are given in the last column of
Table~\ref{tab:cdfd0}.  The CDF measurement has a weight of $60\%$,
while the {\dzero} measurement has a weight of $40\%$. The correlation
between the measurements of the two experiments is 17\%.

\input{table_d0cdfcombi}

\begin{figure*}[htbp]
\includegraphics[width=6.5in]{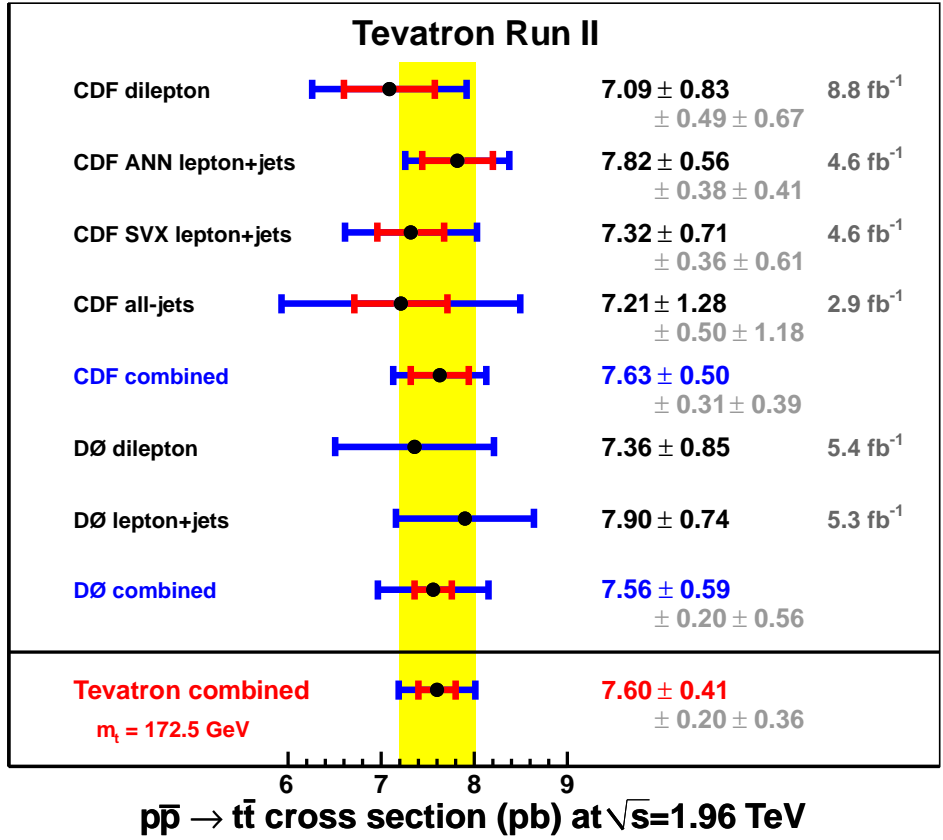}
\caption[xsec-summary]{(color online). The six input
  $\sigma_{t\bar{t}}$ measurements from the CDF and {\dzero}
experiments, along with the CDF-only and {\dzero}-only combination
results, and their combination for the Tevatron result. The total
uncertainty, as well as the statistical and systematic uncertainties
are shown.  The {\dzero} dilepton and \ljets measurements using
constrained nuisance parameters are presented in their published form
indicating only their total uncertainties.  The inner (red) bars
reflect statistical uncertainties while the outer (blue) bars show the
total uncertainties on each measurement. }
\label{xsec-summary}
\end{figure*}

The measurements and the Tevatron combination are shown in
Figure~\ref{xsec-summary}, as well as the results of the CDF-only and
{\dzero}-only combinations.  The Tevatron combination has a $\chi^2$
of 0.01 for one degree of freedom, corresponding to a probability of
92\% to have a less consistent set of measurements.  

The measured $\sigma_{t\bar{t}}$ depends on the value of $m_t$
assumed in the simulation.  For larger values of $m_t$, leptons and jets from
top-quark decay are more energetic and central, and thus more likely
to meet the selection requirements on \pt. The $b$ jets
are also more likely to be identified since $b$-jet identification
efficiency tends to increase with increasing jet \pt.  Both effects
cause the \ttbar\ selection efficiency to increase asymptotically to
its maximum value as the value of $m_t$ assumed in the simulation
increases.  The consequence, for a given data sample, is that the
measured $\sigma_{t\bar{t}}$ decreases as the assumed value of $m_t$
increases. The dependence on $m_t$ is enhanced for methods that
exploit the differences in kinematic properties between \ttbar\ and
background, as the discrimination improves as $m_t$ increases. The
consequence, for a given data sample, is that these methods will identify a
smaller \ttbar content, and measure smaller $\sigma_{t\bar{t}}$,
as the assumed value of $m_t$ increases in the simulation used to
describe \ttbar\ kinematic properties.

Therefore, we also measure $\sigma_{t\bar{t}}$\ for several $m_t$
values at which each experiment has simulated \ttbar\ production and
decay. At \dzero, the fit procedure is repeated for each $m_t$ value,
using systematic uncertainties extrapolated from the central $m_t =
172.5$~GeV. At CDF, the DIL and LJ-SVX counting measurements extract
the selection efficiency for each $m_t$ value, and scale
$\sigma_{t\bar{t}}$ by the ratio relative to that from $m_t =
172.5$~GeV. The LJ-ANN and HAD measurements also repeat the fit
procedure for each $m_t$ value.  We repeat the combination process for
CDF and \dzero, and present the results of the Tevatron combination
for $\sigma_{t\bar{t}}$ at three $m_t$ values in Table~\ref{tab:mass}.
Relative to the central value at $m_t$ of 172.5~GeV, the measured
$\sigma_{t\bar{t}}$ increases by 5\% for an assumed $m_t$ of 170~GeV,
and decreases by 3\% for an assumed $m_t$ of 175~GeV.  This non-linear
dependence is due to the slowing of the rate of increase of
the \ttbar\ selection efficiency as $m_t$ increases.  We parametrize
this dependence through the functional form
\begin{equation}
\sigma_{t\bar{t}}= a + b(m_0 - m_t) + c (m_0 - m_t)^2, 
\label{function}
\end{equation}
with $m_0 = 172.5$~GeV and fitted values of $a = 7.60$~pb, $b =
0.126$~pb/GeV and $c= 0.0136$~pb/GeV$^2$.  The parameters for the fit
corresponding to an upward change of one standard deviation in
$\sigma_{t\bar{t}}$ are $a = 8.01$~pb, $b = 0.132$~pb/GeV and $c=
0.0144$~pb/GeV$^2$, and for a downward change of one standard
deviation are $a = 7.19$~pb, $b = 0.120$~pb/GeV and
$c=0.0128$~pb/GeV$^2$.

\begin{figure}[htb]
\includegraphics[width=3.2in,height=3in]{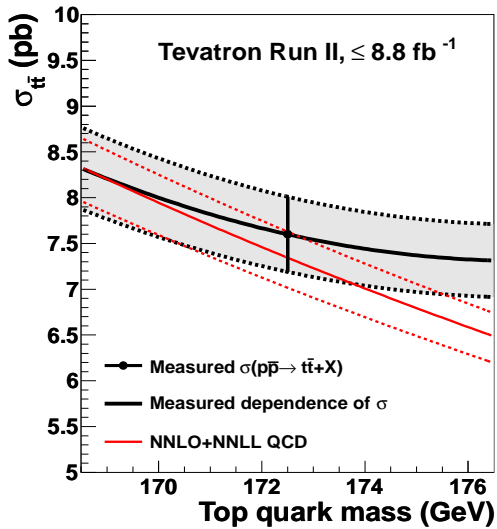}
\caption[tev_xsec_vsmass]{(color online).
The combined $\sigma_{t\bar{t}}$ at the Tevatron as a function of
$m_t$ (black line), as expressed by Eq.(\ref{function}), compared to
the prediction (narrower red line) at NNLO+NNLL in perturbative
QCD~\cite{SMtheory_NNLO}. The dashed lines show the total uncertainty
on the result (black dashed lines enclosing shaded region) and the prediction (narrower red
dashed lines).}
\label{tev_xsec_vsmass} \end{figure}

The dependence of the measured $\sigma_{t\bar{t}}$ on the value for
$m_t$ assumed in the simulation is shown by the shaded band in
Fig.~\ref{tev_xsec_vsmass}.  The measured $\sigma_{t\bar{t}}$ is in
good agreement with the NNLO theoretical prediction for assumed values
of $m_t$ below 175 GeV.

\begin{table}[htb]
\caption{\label{tab:mass} CDF and \dzero\  measurements of $\sigma_{t\bar{t}}$ and their combination, with total uncertainties, for three values of $m_t$.}
\begin{tabular}{lccc}
\hline \hline
Top-quark mass (GeV)	&  170	&  172.5	& 175 \\
\hline
CDF $\sigma_{t\bar{t}}$ (pb) & 8.17 $\pm$ 0.53 &  7.63 $\pm$ 0.50 & 7.35 $\pm$ 0.48 \\
\dzero\ $\sigma_{t\bar{t}}$ (pb) & 7.75 $\pm$ 0.61 & 7.56 $\pm$ 0.59 & 7.40 $\pm$ 0.57 \\ \hline
Tevatron $\sigma_{t\bar{t}}$  (pb) & 8.00 $\pm$ 0.43 & 7.60 $\pm$ 0.41 & 7.37 $\pm$ 0.40 \\ 
\hline \hline
\end{tabular}
\end{table}

\vspace{0.5cm}

%% file: table_d0cdfcombi.tex
\begin{table*}[tbp]
 \caption{CDF and {\dzero} measurements of $\sigma_{t\bar{t}}$ and
   their combination (in pb), with individual contributions to their uncertainties
   (in pb). Correlation indicates whether a given uncertainty
   is treated as fully correlated between the CDF and {\dzero}
   measurements.}
 \label{tab:cdfd0}
\begin{center}
\begin{tabular}{lcccc}
 \hline
 \hline
 & CDF & D0 & & Tevatron \\ 
 \hline
Central value of $\sigma_{\bar{t}t}$ & 7.63 & 7.56 & & 7.60 \\
\hline
Sources of systematic uncertainty & & & Correlation & \\
\hline
Modeling of the detector  & 0.17 & 0.22 & NO & 0.13 \\
Modeling of signal & 0.21 & 0.13 & YES & 0.18 \\
Modeling of jets & 0.21 & 0.11 & NO & 0.13 \\
Method of extracting $\sigma_{t\bar{t}}$  & 0.01 & 0.07 & NO & 0.03 \\
Background modeled from theory & 0.10 & 0.08 & YES & 0.10 \\
Background based on data & 0.08 & 0.06 & NO & 0.05 \\
Normalization of $Z/\gamma^*$ prediction  & 0.13 & -- & NO & 0.08 \\
Luminosity: inelastic $p \bar{p}$ cross section & 0.05 & 0.30 & YES & 0.15 \\
Luminosity: detector & 0.06 & 0.35 & NO & 0.14 \\
\hline
Total systematic uncertainty & 0.39 & 0.56 & & 0.36 \\
\hline
Statistical uncertainty & 0.31 & 0.20 &  & 0.20 \\ 
 \hline
Total uncertainty & 0.50 & 0.59 & & 0.41 \\
 \hline
 \hline
 \end{tabular}
 \end{center}
 \end{table*}

%% file: conclusion.tex
We have presented the combination of measurements of $\sigma_{t\bar{t}}$ in
the dilepton, \ljets, and all-jets final states, using data collected
by the CDF and {\dzero} collaborations at the Tevatron $p\bar{p}$
collider at $\sqrt{s}=1.96$~TeV.  The measurements use data samples
with integrated luminosity between 2.9~fb$^{-1}$ and 8.8~fb$^{-1}$.
Assuming the SM expectation for top-quark decay, we
observe good agreement on $\sigma_{t\bar{t}}$ among the different
experimental final states. The first combination of the CDF and
{\dzero} measurements is 
\begin{equation}
\sigma_{t\overline{t}} = \tevcent \pm \tevtoterr~\mathrm{pb}, \nonumber
\end{equation}
\noindent for a top-quark mass of $m_t=172.5$~GeV.  The combined
$\sigma_{t\bar{t}}$ of $\tevcent$~pb has a relative uncertainty of
$5.4\%$, which is close to the relative uncertainty of the prediction
from theory of about 4\%.  The result is in good agreement with the
latest theoretical expectation for $\sigma_{t\bar{t}}$ in the standard
model, calculated at NNLO+NNLL QCD, of
$7.35^{+0.28}_{-0.33}$\,pb~\cite{SMtheory_NNLO}, as presented in
Fig.~\ref{tev_xsec_vsmass}.

In the future, two improvements to the individual measurements could
reduce the total uncertainty on the combined result by about 25\% to
0.31~pb. Firstly, the {\dzero} measurements and the CDF \ljets channel
measurements could have their statistical uncertainties reduced by a factor
of about 1.4 by updating the published analyses from 5~fb$^{-1}$ to
the full integrated luminosity of about 10~fb$^{-1}$ of data
collected in Run II.  Secondly, the {\dzero} measurements and the CDF
dilepton channel measurement could also reduce their luminosity
uncertainty, as done by the CDF \ljets channel measurements, by
using the ratio of the \ttbar to the $Z/\gamma^*$ cross sections
measured concurrently and then multiplying by the more precisely known
theoretical prediction for the $Z/\gamma^*$ cross section. This
strategy would reduce the current 6\% luminosity uncertainty to a 2\% systematic
uncertainty on the normalization of the $Z/\gamma^*$ prediction.

%% file: acknowledgement.tex
We thank the Fermilab staff and technical staffs of the participating institutions for their vital contributions.
We acknowledge support from the
DOE and NSF (USA),
ARC (Australia),
CNPq, FAPERJ, FAPESP and FUNDUNESP (Brazil),
NSERC (Canada),
NSC, CAS and CNSF (China),
Colciencias (Colombia),
MSMT and GACR (Czech Republic),
the Academy of Finland,
CEA and CNRS/IN2P3 (France),
BMBF and DFG (Germany),
DAE and DST (India),
SFI (Ireland),
INFN (Italy),
MEXT (Japan),
the Korean World Class University Program and NRF (Korea),
CONACyT (Mexico),
FOM (Netherlands),
MON, NRC KI and RFBR (Russia),
the Slovak R\&D Agency,
the Ministerio de Ciencia e Innovaci\'{o}n, and Programa Consolider-Ingenio 2010 (Spain),
The Swedish Research Council (Sweden),
SNSF (Switzerland),
STFC and the Royal Society (United Kingdom),
the A.P. Sloan Foundation (USA),
and the EU community Marie Curie Fellowship contract 302103.

%% file: prd.bbl
\begin{thebibliography}{99}
\bibitem{conjugate}
Here $W^{+}$ also
  refers to the charge conjugate state with the fermion charges
  conjugated as well. In the following the
  charges of the $W$ boson and the leptons $\ell$ are dropped.
\bibitem{DeltaR} $\cal R$ is defined as ${\cal R}=\sqrt{ (\Delta
    \eta)^2 + (\Delta \phi)^2 }$.  The pseudorapidity $\eta$ is a
    function of the polar angle $\theta$ relative to the proton beam:
    $\eta(\theta) = - \ln [ \tan (\theta/2)]$. We distinguish detector
    $\eta$ ($\eta_{det}$) and collision $\eta$, where the former is
    defined with respect to the center of the detector and the latter
    relative to the $p\bar{p}$ interaction vertex.  The angle $\phi$ is the
    azimuthal angle around the beamline.
\end{thebibliography}
